\documentclass[runningheads,citeauthoryear]{apinv}
\usepackage{epsfig,cite,graphics}
\usepackage{epsfig,cite,graphics}
\usepackage[T2A]{fontenc}
\usepackage[cp1251]{inputenc}
\usepackage[bulgarian,english]{babel}
\begin{document}
\setcounter{page}{1}
\title{Intra-night flickering of RS Ophiuchi: III.\\
Modes of quasi-periods in the minute scale\\ and their evolution}
\author{Ts. B. Georgiev\inst{1},
R. K. Zamanov\inst{1}, 
S. Boeva\inst{1}, 
G. Latev\inst{1},
B. Spassov\inst{1}, \\  
J. Mart\'{\i}\inst{2}, 
G. Nikolov\inst{1}, 
S. Ibryamov\inst{3},  
S. V. Tsvetkova\inst{1},  
K. A. Stoyanov\inst{1}}
\titlerunning{ Flickering of RS~Oph. III.}
\authorrunning{Ts. B. Georgiev, R. K. Zamanov, S. Boeva et al.}
\tocauthor{Ts. B. Georgiev,
R. K. Zamanov, 
S. Boeva, 
G. Y. Latev,
B. Spassov, 
J. Mart\'{\i}, 
G. Nikolov, 
S. Ibryamov,  
S. V. Tsvetkova,  
K. A. Stoyanov }

\institute {Institute of Astronomy and National Astronomical Observatory, \\ Bulgarian Academy of Sciences, 72 Tsarigradsko Chaussee Blvd., 
	  1784 Sofia, Bulgaria
	\and Departamento de F\'{\i}sika (EPSJ), Universidad de Ja\'{e}n,\\ Campus Las Lagunillas  A3-420, 23071 Ja\'{e}n, Spain
	\and University of Shumen, 115, Universitetska Str., 9700 Shumen, Bulgaria
\newline \email{tsgeorg@astro.bas.bg }}
\papertype{Submitted on XX.XX.2019; Accepted on XX.XX.2019}	
\maketitle

\begin{abstract}
We study the photometric behavior of the recurrent nova RS~Oph by 58 monitoring light curves (LCs), taken by 5 telescopes. All LCs show repeating time structures with some quasi-periods (QPs) in time scales from minutes to hours. In our previous work 97 QPs were detected in the LCs by local minimums of structure functions and 
local maximums of auto-correlation functions. The distribution of the QPs shows modes at 8, 13, 21, 30, 48 and 73 min, where the mode at 8 min is poorly unveiled. These modes follow a power function with base $1.55\approx 3/2$ with standard deviation 4.7\%. This function predicts modes also at 5.3 and 3.5 min, which are not detected in the full MLCs. In the present work we analyze simple small parts from high resolution LCs. We confirm the QPs modes at 8.0, 5.3 and 3.5 min. Generally, we found 8 QP modes with regular logarithmic distribution in the time interval 3.5--73 min. We also show typical intra-night evolutions of QP modes in the minute scale -- sharp or gradual transitions from one QP mode to other. In the end we find that the parts of the LCs carry out the properties of the whole LCs at short time scale. This lead to two well pronounces dependences -- between the range deviation and standard deviation of the LC, as well as between the quasi-period and the relevant level of the density function of the LC.   
\end{abstract}

\keywords{stars: binaries: symbiotic -- novae, cataclysmic variables -- accretion, accretion discs -- stars: individual: RS Oph }
\\

\section*{Introduction}

RS~ Oph is a symbiotic recurrent nova containing an M2~III mass donor (She-navrin, Taranova \& Nadzhip 2011) and a massive carbon-oxygen white dwarf (Mikolajewska \& Shara 2017). The flickering of RS~Oph (brightness variability on time scales from minutes to hours) has been firstly detected by Walker (1977). Wynn (2008) proposed that both Roche lobe overflow and stellar wind capture are possible accretion scenarios in the case of RS~Oph.

Kundra, Hric \& G\'{a}lis (2010) carried out wavelet analysis and unveiled flickering quasi-periods (QPs) with modes at 15--30 min and $> 30$ min. Later, Kundra \& Hric (2014) revealed FQP modes at 10 min and 24 min. 

In our Paper I (Georgiev et al. 2019a), working in magnitude scales, we found repeating flickering structures with QPs of 10--120 min in 
all our 58 monitoring light curves (MLCs), with modes of the dominating QPs at 10, 21, 36 and 73 min. We found also a faint correlation between the QP and its flickering energy, 
estimated by the relevant plateau of the standard deviation function of the residual LC (LC with removed linear trend). 

Zamanov et al. (2018) derived the apparent magnitudes of the red giant, $m_B$=14.66 mag and $m_V=12.26$ mag, as well as interstellar extinction toward RS~Oph, $E(B-V)=0.69$. 
%The relevant magnitude corrections are $A(B)=4.19\times E(B-V)$ and $A(V)=3.16\times E(B-V)$. 
Thus ensures the study of the behavior of the flickering source (FS) in the linear scale of fluxes. Zamanov et al. (2018) found that RS Oph becomes bluer as it becomes brighter; however the hot component becomes redder as it becomes brighter. No correlation between the temperature of the FS and the brightness of FS was found. However, a strong correlation (correlation coefficient 0.81) between the $B$ band magnitude and the average radius of the flickering source was revealed; that is, as the brightness of the system increases, the size of the flickering source also increases. Zamanov et al. (2018) found also a dependence between the $B$ band magnitude and the radius of the FS with correlation coefficient 0.81. 

In our Paper II (Georgiev et al. 2019b) we worked in linear scales of fluxes. The large scale trends of the FLCs were removed by suitable fitting polynomials. Specific numerical techniques, called also fractal analyzes, were applied. Some parameters and functions for quantitative characterizing of the MLCs were established. Further 97 QPs with sizes from 6 to 80 min were found. The distribution of the QP modes, found in Paper II, more accurate than in Paper I, occur again not arbitrary.

Figures 12 in Paper II visualizes the regularity (near-commensurability) of the QP modes distribution. The observed modes are placed at about 8, 13, 21, 30, 48 and 73 min. The 8 min mode is poorly  pronounced and we regarded it as suspected. The power function of the QP $P$ on the serial number of the mode $M$ is 
 \begin{equation}     % Eq. 1
 P_M=3.47 \times 1.55^M \pm 4.7\% 
 \hspace{4mm} \mathrm{or} \hspace{4mm} 
 \lg P_M = 0.54 + 0.19\times M \pm 0.02.
\end{equation} 
Note the small standard deviations (SDs) of these dependencies. The respective graphs, supplemented with the result of the present work   (Sect.~2), are shown in Fig.~3(a,b). 

The power function (Eq.~1) predicts QP modes also at 3.5, 5.3 and 73 min, not detected in Paper II. The number of the suspected QP mode at 3.5 min is accepted to be $M=0$. Short QPs obviously exist in the LCs,  taken with high resolution. The characteristic duration of the elementary shot of $3.6\pm 0.8$ min (Paper II, Fig.~11) corresponds to the predicted QP mode $M=0$ at 3.5 min.

Today the flickering phenomenon is explained qualitatively by variable mass transfer from mass donor through the accretion disk to 
the surface of white dwarf (e.g. Lyubarskii 1997). The reasons of the of the different QP modes and their time evolution are not clear. 

The purposes of this paper are (i) to check whether the flickering in  the minute scale supports the predicted in Paper II QP modes at 8.0, 5.3, 3.5 min, as well as (ii) to show how much the shortest QPs evolve in the bounds of one LC. The contents follows:\\
1. Observing material and data reduction; \\
2. Confirmation of the 8.0, 5.3 and 3.5 min QP modes; \\
3. Intra-night evolutions of the minute QPs modes; \\
4. The Parts of LC and the whole LCs; \\
   Conclusions;  Bibliography;  Appendix A and B.

Appendix A contains Table A1 with data about the parts of the LCs and Table A2 with data about the unveiled QPs. Appendix B contains 8 panels with illustrations, similar to Fig.~1. 

	Abbreviations: \\ 
ACF -- auto-correlation function (Eq.~4, Fig.~1(d)); \\
FS -- flickering source;  \\
LC -- monitoring light curve (Fig.~1(a)); \\
QP -- quasi-period.\\
SD -- standard deviation;\\
SF -- structure function (Eq.~3, Fig.~1(c)).\\

\section*{1. Observing material and data reduction}

We explore 58 linearized FLCs of the FS (28 in B band, 29 in V band), taken in 2009--2018 by 5 telescopes. The apparent magnitude of RS~Oph is 10--11 mag. The standard error of the photometry is 0.005--0.010 mag. The observing material is described in the paper of Zamanov et al. (2018) and in Paper II. We are working in the linear flux scale. The details of the processing are described in Paper II. 

For detection QPs in the minute scale we selected 9 parts of FLCs with high time resolution and without single strong fluctuations (Fig.~1, Appendix A, B). The sizes of the FLC parts are 18--46 min, the time steps are  0.24--0.70 min and the auto-correlation times are 0.6--1.7 min. 

%\newpage \vspace{-4mm}
\begin{figure*}[!htb]  \begin{center}
\centering{\epsfig{file=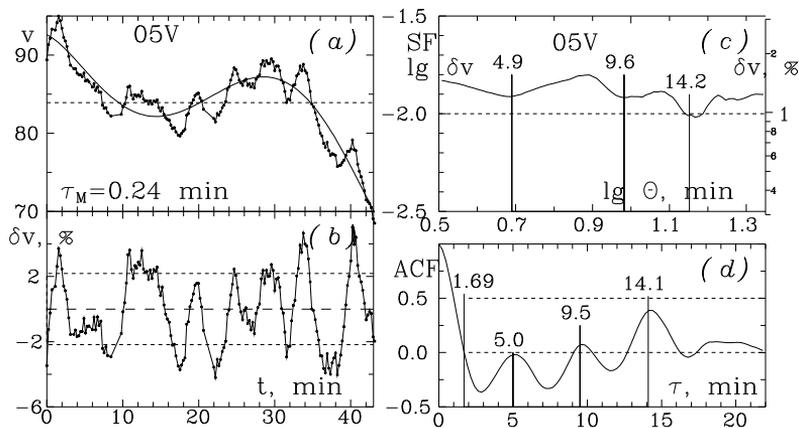, width=.8\textwidth}}
%\vspace{-4mm}
\caption[] {Processing of part of the LC \#05V (see the text). (\textbf{a}): The linearized LC with average time step $\tau_M = 0.24$ min, 5-th degree fitting polynomial and the average level; 
(\textbf{b}): The relevant residual relative LC, in percents, and its $\pm$~SD interval; 
(\textbf{c}): The plateau of the structure function, SF, of LC in (b) and its 1\% level. The right ordinate is marked in percents. The local minimums indicate QPs, marked by vertical segments; 
(\textbf{d}): The auto-correlation function, ACF, of the LC in (b) and its 0.0 and 0.5 levels. The local maximums indicate QPs, marked by vertical segments. The leftmost number is the auto-correlation time,  1.69 min. }
\label{fig.1}   \end{center}  \end{figure*} % \vspace{-6mm}

Figure 1 (a) presents the processing stages of the selected part of the LC \#05V. These parts of LC is transformed from magnitudes to flux scales in units $10^{-9}$~erg~ cm$^{-2}$~s$^{-1}$~\AA$^{-1}$ as in Paper II. Figure 1(b) presents the relevant residual relative LC:   
 \begin{equation}     % Eq. 2
 \delta v(t) = [v(t)-v^{(5)}(t)]/ v^{(5)}(t), 
\end{equation} 
where $v^{(5)}(t)$ is 5-th degree fitting polynomial, shown in (a). In $B$ band analogous formula is valid. For easy  compatibility the LCs $\delta v(t)$ or $\delta b(t)$ in sub-figures (b) are expressed in percents. 

Figures 1(c) shows the structure function (SF) of the LC  in (b):  
\begin{equation}    %Eq. 3
SF = \mathrm{lg} <v(t+\theta)-v(t)> = F(\lg\theta),  
\end{equation}
where $\theta$ is the time lag. The broken brackets mean averaging for all positions of $\theta$ in the LC. The local minimums detect QPs. In a quasi-periodic time series the use of the SF as detector of QPs is equivalent to the "phase dispersion minimization technique" of  Lafler \& Kinman (1965).
 
Figures 1(d) shows the auto-correlation function (ACF) of the residual LC in (b):   
\begin{equation}    %Eq. 4
ACF = <v(t+\tau)\times  v(t)> = F(\tau), 
\end{equation} 
where $\tau$ is the time lag of the ACF. The local maximums detect QPs.

Both SF and ACF functions are applicable on LCs only after removal of the large scale trends. This is done in the residual LCs (Eq.~2). We note that only for the ACF the linearized MLCs are re-sampled with the average time step $\tau_M$ of the used part of the LC. The same transform is applied for the ACFs on the whole MLCs in Paper II.

The Appendix B shows other 8 parts of FLCs, treated by the same manner, where the four diagrams, shown in Fig.~1, are ranged vertically. The fitting polynomial for the removal of the large scale trend, is always of 5-th degree. The QPs are found separately by SFs and ACFs are used together, in a common list.

\section*{2. Confirmation of the 8.0, 5.3 and 3.5 min QP modes}

Figure 2 shows the distributions of 63 short QPs, found in this work,  30 from SFs and 33 from ACFs. The QP modes at 8.0, 5.3 and 3.5 min are confirmed. Only the QP mode at 13 min, observed well in Paper II, seems to be dual. The reason is the shortness of the used parts of the MLCs. However, over the logarithmic time axis (Fig.~3) the QP at 13 min is well unveiled. 

%\newpage \vspace{-4mm}
\begin{figure*}[!htb]  \begin{center}
\centering{\epsfig{file=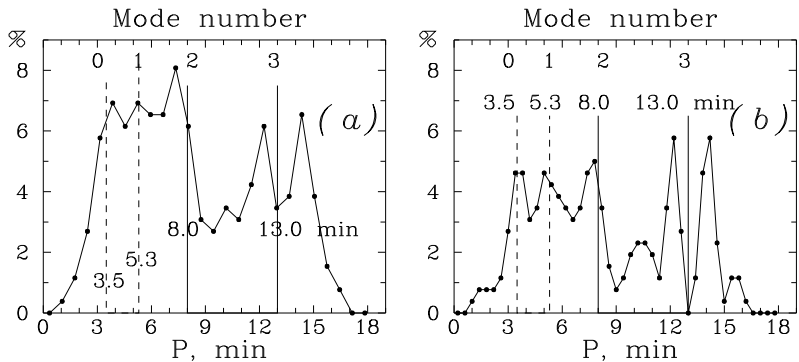, width=.8\textwidth}}
%\vspace{-4mm}
\caption[] {Confirmation of the QP modes at 8.0, 5.3 and 3.5 min, suspected in Paper II. The distribution of 63 short quasi-periods, found in this work, is presented by binning intervals of 0.7 min (\textbf{a}) and 0.4 min (\textbf{b}), respectively.
The histograms in Figs.~2 and 3 are slightly smoothed by convolution kernel with coefficients [0.25, 0.50, 0.25]. }
\label{fig.2}   \end{center}  \end{figure*} % \vspace{-6mm} 

%\newpage \vspace{-4mm}
\begin{figure*}[!htb]  \begin{center}
\centering{\epsfig{file=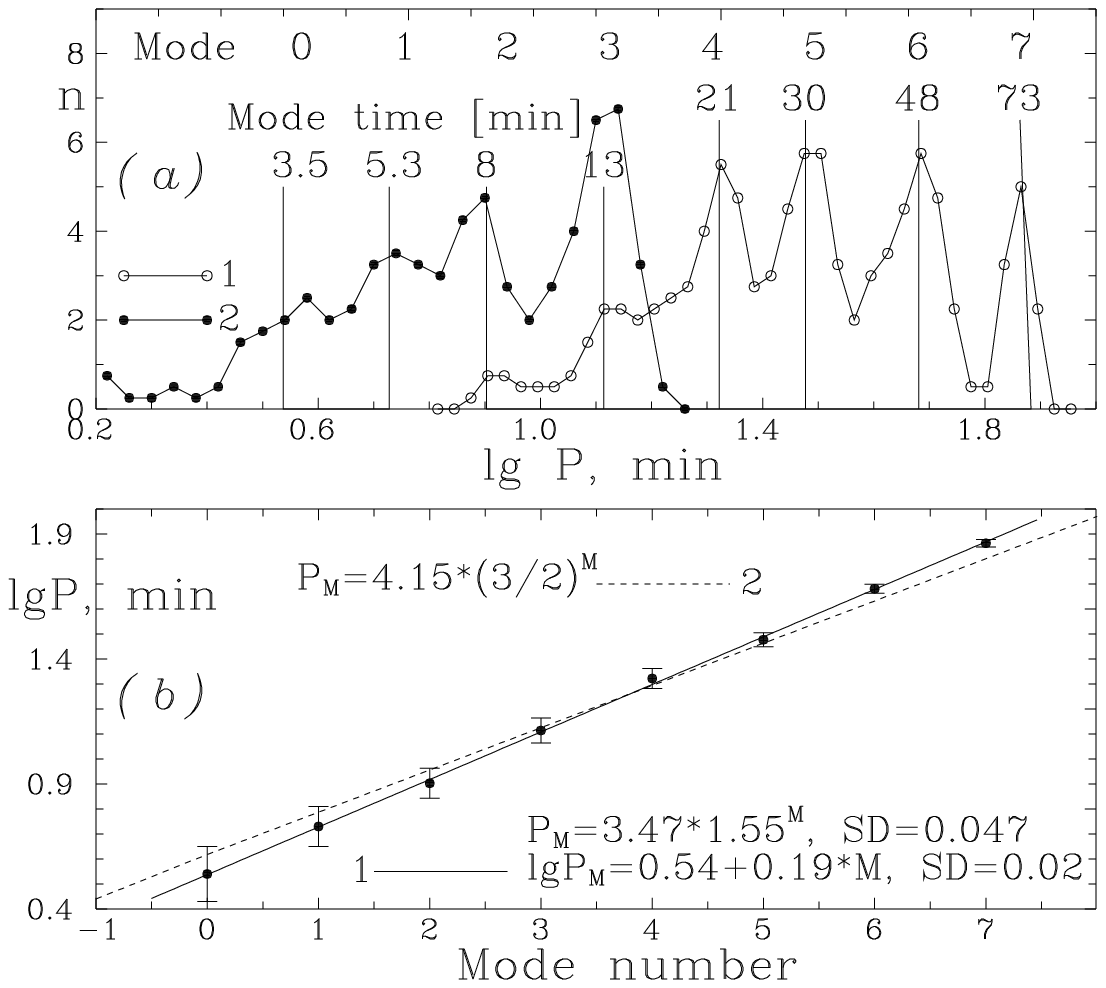, width=.75\textwidth}}
%\vspace{-4mm}
\caption[] { (\textbf{a}): Joint distribution of 160 QPs found in Paper II (line 1) and in this paper (line 2) over logarithmic scale of the time. The binning steps $\Delta\lg t$ of the distributions are 0.025 and 0.040, respectively. Th vertical segments mark the QP modes. 
(\textbf{b}): The regularity of 8 QP modes along the power function (solid line, bottom formulas). The close power function (dashed line, top formula) with base 1.5=3/2, is build though the mode of 21 min.}
\label{fig.3}   \end{center}  \end{figure*} % \vspace{-6mm}

Figure 3 shows the main result form Paper II and this paper -- regular  distribution of 8 QP modes in the time interval 3.5--73 min, following power function with base 1.55 (Eq.~1). The modes at 3.5 and 5.3 min are poorly resolved probably by the evolution of the short QPs in the  bounds of one monitoring (Sect.~3).  

A close power function, with base 3/2, $P_M=4.15\times 1.5^M$, is shown in Fig.3(b) too. The difference between the bases of both power functions, of about 0.05, overcome weakly the 90\% Student's criterion of confidence. It seems these small difference is a hint about somewhat 3/2 resonance sequence of the flickering modes.    

%\newpage \vspace{-4mm}
\begin{figure*}[!htb]  \begin{center}
\centering{\epsfig{file=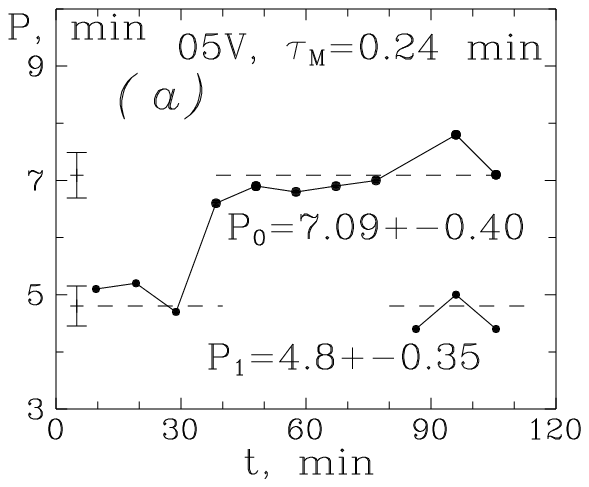, width=.4\textwidth}}
\centering{\epsfig{file=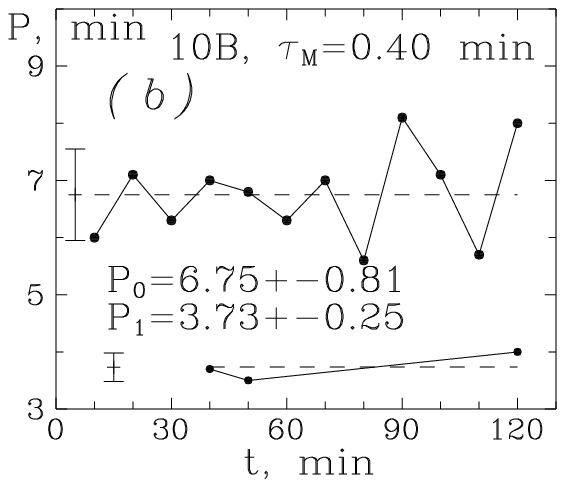, width=.4\textwidth}}
\centering{\epsfig{file=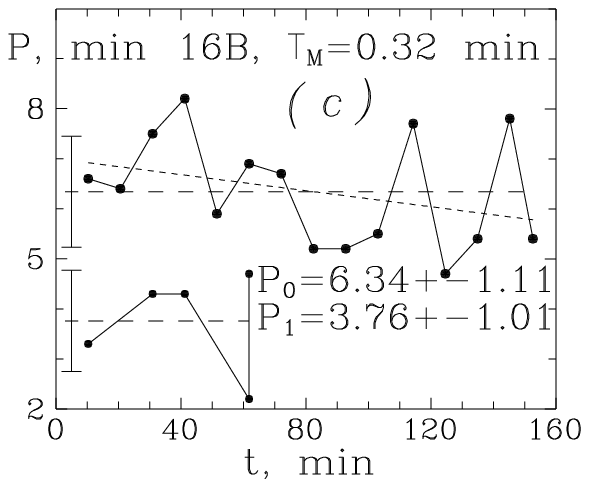, width=.4\textwidth}}
\centering{\epsfig{file=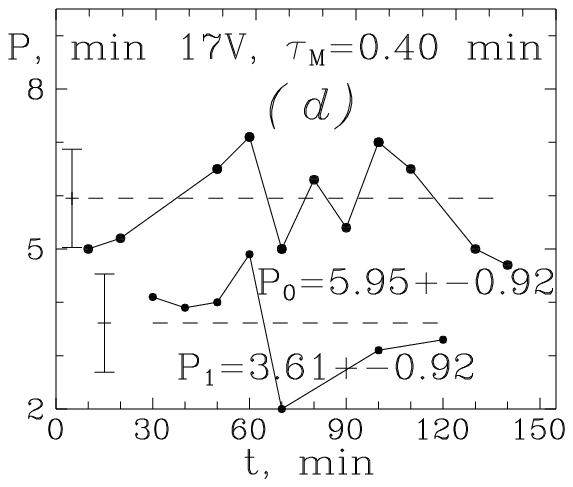, width=.4\textwidth}}
%\vspace{-4mm}
\caption[] {Typical evolutions of short time QP modes in the minute scale. Top and bottom broken lines correspond to the "main" and "alternative" QPs, respectively. The average values and the SD of the main and alternative QPs, $P_0$ and $P_1$, respectively, as well the time step $\tau_M$ of the LC, are implemented. (\textbf{a}): Jump from shorter to longer mode: (\textbf{b}): Oscillations about some mode; (\textbf{c}) Oscillations with decreasing of the mean mode; (\textbf{d}): Evolution from short though long and again to short mode.}
\label{fig.4}   \end{center}  \end{figure*} % \vspace{-6mm} 

\section*{3. Intra-night evolutions of the minute QPs modes}

In Papers I and II we detected QPs with durations from minutes to hours. Undoubtedly, these modes cannot be constants for long time. By this reason we check the time evolution of the shortest QPs modes. We selected only 4 suitable LCs -- LCs with long duration and high resolution. To follow the evolutions of the minute modes we move a probe time window with size 19--20 min and step 1/2 of the size. For each fixed position of the window we derive main QP and, eventually, alternative QP. 

Figure 4 shows the results. In these cases the average QP modes occur between 5.95 and 7.09 min, just between the QP modes of 5.3 and 8 min. The ratio between the average values of the main QPs and alternative QPs is $P_0/P_1=1.66\pm 0.13$, which again seems near to 3/2. But the statistics here is poor. 

\section*{4. The parts of the LCs and the whole LCs}

The selected 9 parts of LCs may be regarded as separate system of LCs. Below we compare these parts with the system of the whole LCs, analyzed in Paper II. The data about the parts of LCs and results of their processing are collected in Tables in the Appendix A.

In correspondence with Fig.~3 in Paper II, the parts of LCs pose   intermediate asymmetry (skewness, $0.08\pm 0.23$, from --0.22 to 0.55) and excess (kurtosis, $-0.09 \pm 0.65$, from --0.63 to 1.29,  after removal of an extraordinary high kurtosis). The asymmetry and excess of the parts of the LCs are very similar to these for the whole LCs. They correlate as in Fig.~3(c) in Paper II. 

The structure gradients (as it is defined in Paper II, Sect.~2.3) of the regarded small LC parts are poorly pronounced, close to zero and useless. They are not shown in sub-figures (c) in Fig.~1 and in Appendix B. However, the Hurst gradients (Paper II, Sect.~2.4) are well pronounced, $0.28 \pm 0.05$, taking again a narrow interval of values, from 0.22 to 0.37 (in this case the extraordinary low value 0.13 of \#05V, shown Fig.~1, is removed). The relevant average fractal dimension is $2 - 0.28 - 0.3 = 1.42 \pm 0.05$. It is close to the respective value of $1.48 \pm 0.06$ in Paper II, Fig.~8(b).

The processed parts of LCs show correlations between the average flux and the standard deviation or the range deviation, as in Paper II, Fig.~2(b) but with low slope coefficients. The reason is the parts of the LC are chosen to be simple (with not large fluctuations). However, these "quiet" parts of LC lead to two dependences -- "morphologic" and "energetic". They seem to be useful for classifications of flickering and similar time series. 

Figure 5(a) shows dependence of the range deviation (RD, half of the peak-to-peak amplitude) on the standard deviation (SD) of the LC ($\delta b$ or $\delta v$). The implemented parameters are: correlation coefficient (CC), gradient (R, slope coefficient) and standard deviation (SD). The regression gradient $GR=2.45$, is derived from 67 data (58 LCs from Paper II, circles, and 9 LCs from this work, dots. Dashed lines show the relevant such gradients of 4 other time series, noted in Paper II: W -- Wolf index of the sunspot number with $GR=3.11$, V -- number of visits in the Smolyan Planetarium with $GR=2.27$, U -- uniform random process, $GR=1.72$ and G -- Gaussian random process with $GR=3.11$. The flickering of RS~Oph takes well defined intermediate place.           

%\newpage \vspace{-4mm}
\begin{figure*}[!htb]  \begin{center}
\centering{\epsfig{file=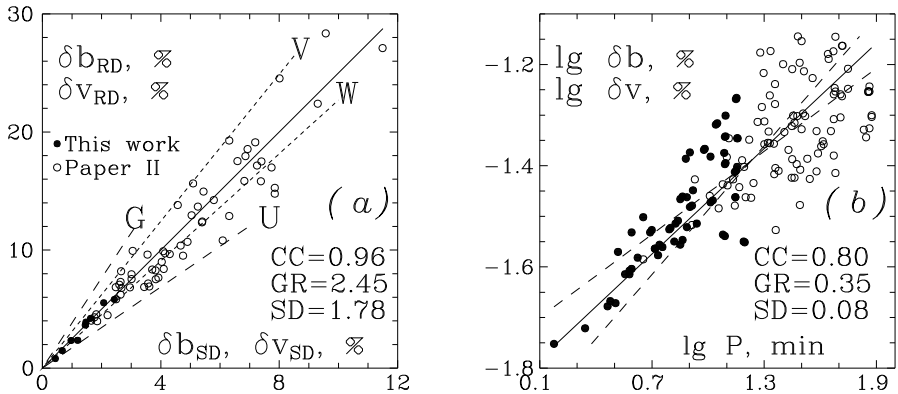, width=.9\textwidth}}
%\vspace{-4mm}
\caption[] {Two regression dependences with implemented parameters correlation coefficient (CC), gradient (GR) and standard deviation (SD). (\textbf{a}): Proportionality between RD and SD (solid line) for 67 LCs. The dashed lines visualize the relevant gradients of 4 other time series (see the text);    
(\textbf{b}): Proportionality between the QP, $P$ and the level of the density function by 168 data, in log-log coordinates. The dashed libes show the ordinary and reverse regression and the solid line show their bissectrice. }
\label{fig.5}   \end{center}  \end{figure*} % \vspace{-6mm} 

Figure 5(b) presents the dependence of the density function (Paper II, Sect.~2.2) on the quasi-period $P$ by 168 QP (94 from Paper II and 64 from this work). The amplitudes of the large QPs are very different and by this reason toward the large QPs the width of this dependence enlarge. 

So, the parts of the LCs carry out the properties of the whole LCs. The dependences shown in Fig.~5 seems to be fundamental and may be basic tools for comparison of flickering processes.

The parts of the LCs again do not answer clearly to the question about the dominating feedback of the flickering -- local shots of global disk instabilities. Obviously, the flickering phenomenon is complicated in all its manifestations. 

\section*{Conclusions}

In this paper we analyze 9 parts of LCs (Eq.~2), carried out with  observing resolution 0.24--0.70 min with average fractal dimension $1.42 \pm 0.05$. 

1. We confirm the QP modes at 8.0, 5.3 and 3.5 min, predicted in Paper II. Generally, 8 QP modes obey regularity (near-commensurability) corresponding to a power function with base $1.55 \approx 3/2$ (Fig.~3).

2. We also represent typical evolutions of the QP modes in the minute scale, including jumps from one mean FQP to other (Fig.~4). 
We consider analogous changes of longer QP modes exist also on time scales of ours. 

3. We find that the parts of the LCs carry out the properties of the whole LCs. The "morphologic" and "energetic" dependences, shown in Fig.~5 seems to be fundamental and may be basic tools for comparison of flickering processes.

In the next paper we intend to present the typical shapes of the repeating flickering structures.  

{\bf Acknowledgments: }
This work is supported by the grant K$\Pi$-06-H28/2 08.12.2018
(Bulgarian National Science Fund).

The authors are grateful to the anonymous referee for its attention to this paper and for the recommendations. 

{}     

\begin{appendix}
\section*{Appendix A. Table A1.}   
\begin{table}

\centering
\caption{\textbf{Data about the processed parts of flickering light curves (LCs).} 1~-- designation of the LC, 2~-- monitoring duration of the LC (min),  3~-- number of the points in the LC, 4~-- mean time step of the LC (min), 5~-- average flux of the LC (in $B$ or $V$ band), 6~-- standard deviation of the LC, 7~-- half-range (amplitude) deviation of the LC, 8~-- asymmetry (skewness) of the deviations in the LC, 9~-- excess (kurtosis) of the deviations in the LC, 10~-- year of the observation.}
\begin{tabular}{cccccccccc}
\hline\noalign{\smallskip}
\# & $T_M$ & $N_M$ & $\tau_M$ & $D_\mathrm{AV}$ & $F_\mathrm{SD}$  &  $F_\mathrm{RD}$ & $As$ & $Ex$ & Year \\
\hline\noalign{\smallskip}
1 & 2 & 3 & 4 & 5 & 6 & 7 & 8 & 9 & 10  \\
\hline\noalign{\smallskip}
p10b &  18 &  45 &  0.40 &   68.7 &  1.19 &  2.36&  --0.048 & --0.825 &   2012.5503 \\
p13b &  28 &  48 &  0.60 &   69.4 &  0.97 &  2.34& ~\,0.223 &~\,0.272 &   2012.6297 \\
p16b &  24 &  76 &  0.32 &  131.0 &  2.43 &  5.82& ~\,0.140 & --0.348 &   2013.6188 \\
p17b &  30 &  75 &  0.40 &   77.8 &  1.45 &  3.65&  --0.272 &~\,0.201 &   2013.6215 \\
p02v &  20 &  30 &  0.70 &  110.9 &  1.63 &  4.21& ~\,0.113 &~\,0.916 &   2009.4545 \\
p05v &  44 & 186 &  0.24 &   83.9 &  2.07 &  5.54& ~\,0.141 & --0.623 &   2009.5640 \\
p13v &  15 &  32 &  0.48 &   32.5 &  0.44 &  0.81& ~\,0.016 & --0.557 &   2012.6297 \\
p16v &  23 &  53 &  0.45 &   62.5 &  1.47 &  3.90& ~\,0.552 &~\,1.294 &   2013.6188 \\
p17v &  17 &  47 &  0.38 &   37.4 &  0.67 &  1.49&  --0.184 & --0.247 &   2013.6215 \\
\hline
\end{tabular}   \end{table}
   
\newpage
\section*{Appendix A. Table A2.}    
\begin{table}   \centering
\caption{ \textbf{Results about the short quasi-periods.} 1, 4, 7~-- designation of the LC, 2, 5, 8~-- quasi-period in minutes, 3, 6, 9~-- respective value of the density function in $B$ or $V$ band in fluxes (Paper II).}
\begin{tabular}{cccc|cccc|cccc}
\hline\noalign{\smallskip}      
\# & $P$ & $\delta b$ or $\delta v$ & &
\# & $P$ & $\delta b$ or $\delta v$ & & 
\# & $P$ & $\delta b$ or $\delta v$ & \\
\hline\noalign{\smallskip}   
1 &   2 &  3 & &  4  &  5  &  6  & &  7   &   8   & 9  \\
\hline\noalign{\smallskip} 
10B  &  4.5  &  0.0260 & &   17B &    5.4 &   0.0265 & &   13V &    4.2  &  0.0262 \\
10B  &  6.3  &  0.0300 & &   17B &   12.3 &   0.0289 & &   13V &    7.3  &  0.0346 \\
10B  &  8.0  &  0.0330 & &   17B &   15.7 &   0.0281 & &   13V &   12.3  &  0.0454 \\
10B  & 14.3  &  0.0396 & &   02V &    2.9 &   0.0210 & &   16v &    3.8  &  0.0247 \\
10B  &  3.8  &  0.0243 & &   02V &    5.7 &   0.0275 & &   16v &    7.7  &  0.0345 \\
10B  &  6.2  &  0.0298 & &   02V &    8.7 &   0.0306 & &   16v &   14.4  &  0.0451 \\ 
10b  &  8.2  &  0.0332 & &   02V &   14.0 &   0.0345 & &   16v &    2.2  &  0.0190 \\
13B  &  7.3  &  0.0284 & &   02V &    3.2 &   0.0213 & &   16v &    3.9  &  0.0249 \\
13B  & 10.3  &  0.0337 & &   02V &    5.5 &   0.0278 & &   16v &    8.3  &  0.0356 \\
13B  & 14.1  &  0.0389 & &   02V &    7.7 &   0.0301 & &   16v &   12.2  &  0.0422 \\
13B  &  7.1  &  0.0278 & &   02V &   14.0 &   0.0345 & &   16v &   14.3  &  0.0451 \\
13B  & 10.6  &  0.0340 & &   05V &    5.0 &   0.0297 & &   17v &    1.5  &  0.0177 \\
13B  & 13.9  &  0.0387 & &   05V &    9.5 &   0.0428 & &   17v &    3.3  &  0.0269 \\
16B  &  5.2  &  0.0273 & &   05V &   14.1 &   0.0539 & &   17v &    4.5  &  0.0315 \\
16B  &  6,9  &  0.0310 & &   05V &    4.9 &   0.0294 & &   17v &    8.0  &  0.0423 \\
16B  & 12.4  &  0.0403 & &   05V &    9.6 &   0.0430 & &   17v &   11.0  &  0.0480 \\
16B  &  5.2  &  0.0273 & &   05V &   14.2 &   0.0542 & &   17v &    3.9  &  0.0294 \\
16B  &  6.7  &  0.0306 & &   13V &    3.6 &   0.0243 & &   17v &    7.6  &  0.0411 \\
16B  & 12.3  &  0.0401 & &   13V &    7.1 &   0.0342 & &   17v &   11.2  &  0.0483 \\
17B  &  6.6  &  0.0282 & &   13V &   10.3 &   0.0415 & &   17v &   12.4  &  0.0500 \\
17B  & 12.0  &  0.0291 & &   13V &   12.4 &   0.0455 & &   --  &     --  &   --    \\ 
17B  &  15.5 &  0.0282 & &   13V &    3.0 &   0.0215 & &   --  &     --  &   --    \\
\hline\noalign{\smallskip}
\end{tabular}  \end{table}
   
\newpage   
\section*{Appendix B. Parts of MLCs \#10B and \#02V. See Fig.~1. }
\begin{figure*}[!htb]  \begin{center}
\centering{\epsfig{file=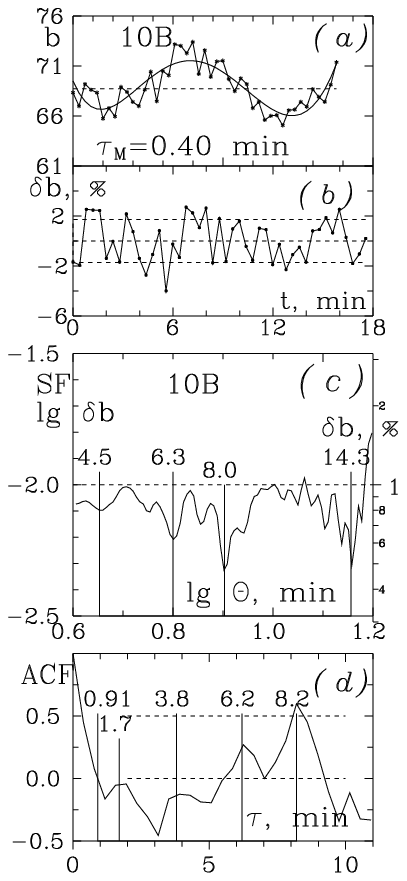, width=.49\textwidth}}
\centering{\epsfig{file=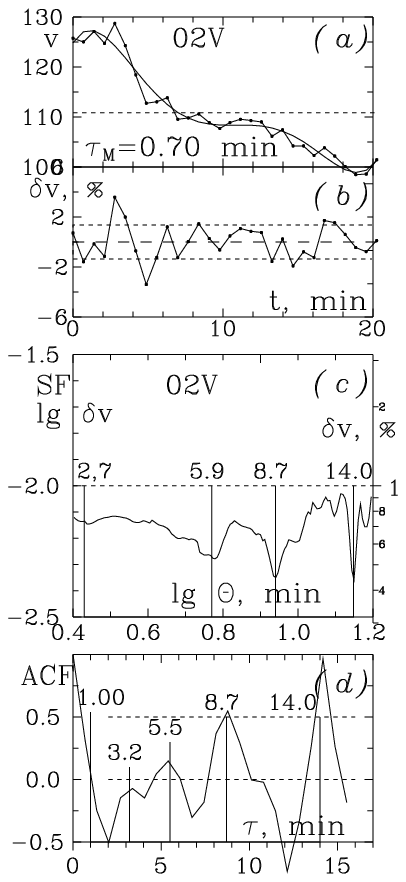, width=.49\textwidth}}
%\caption[] { }
\label{fig.2}   \end{center}  \end{figure*} % \vspace{-6mm}

\newpage
\section*{Appendix B. Parts of MLCs \#13B and \#13V. See Fig.~1. }
\begin{figure*}[!htb]  \begin{center}
\centering{\epsfig{file=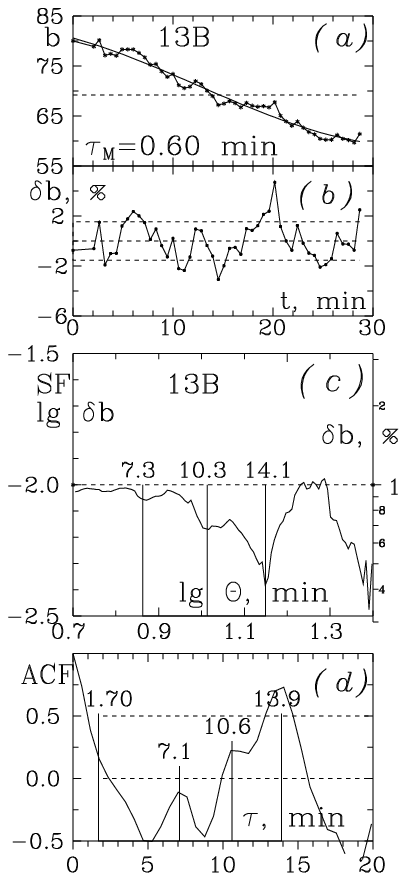, width=.49\textwidth}}
\centering{\epsfig{file=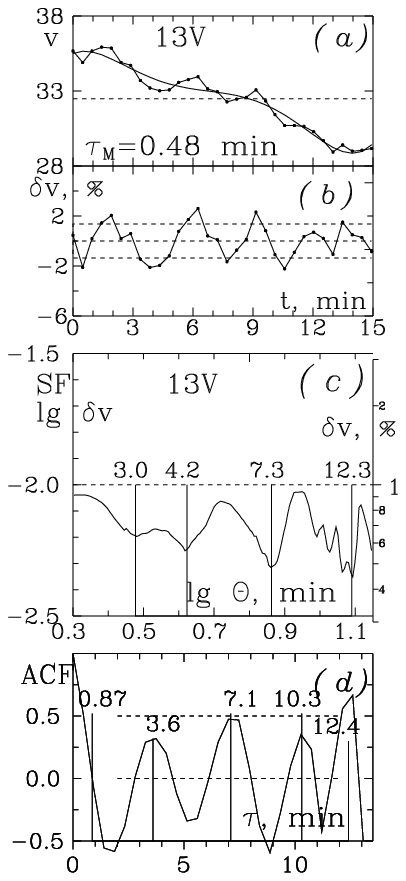, width=.49\textwidth}}
%\caption[] { }
\label{fig.2}   \end{center}  \end{figure*} % \vspace{-6mm}

\newpage
\section*{Appendix B. Parts of MLCs \#16B and \#16V. See Fig.~1. }
\begin{figure*}[!htb]  \begin{center}
\centering{\epsfig{file=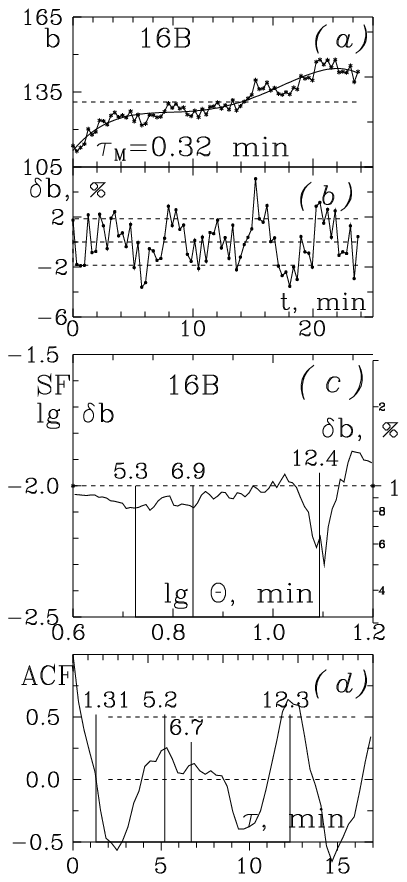, width=.49\textwidth}}
\centering{\epsfig{file=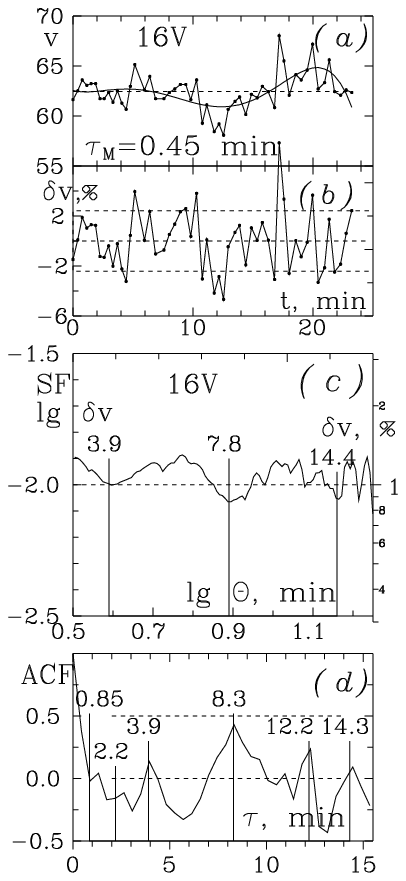, width=.49\textwidth}}
%\caption[] { }
\label{fig.2}   \end{center}  \end{figure*} % \vspace{-6mm}

\newpage
\section*{Appendix B. Parts of MLCs \#17B and \#17V. See Fig.~1. }
\begin{figure*}[!htb]  \begin{center}
\centering{\epsfig{file=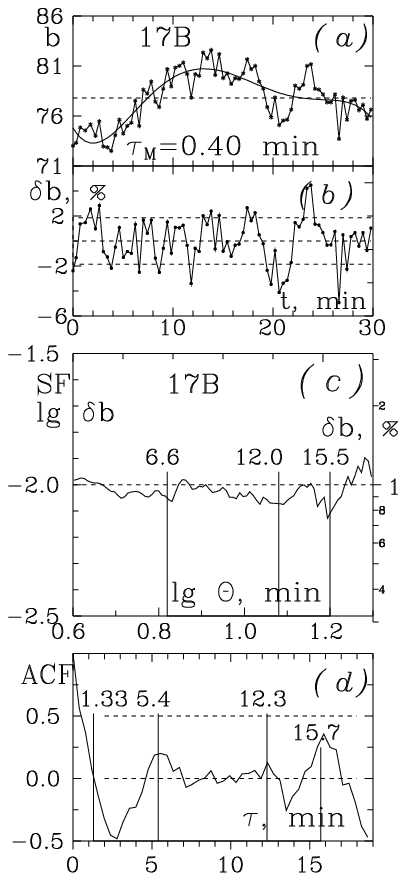, width=.49\textwidth}}
\centering{\epsfig{file=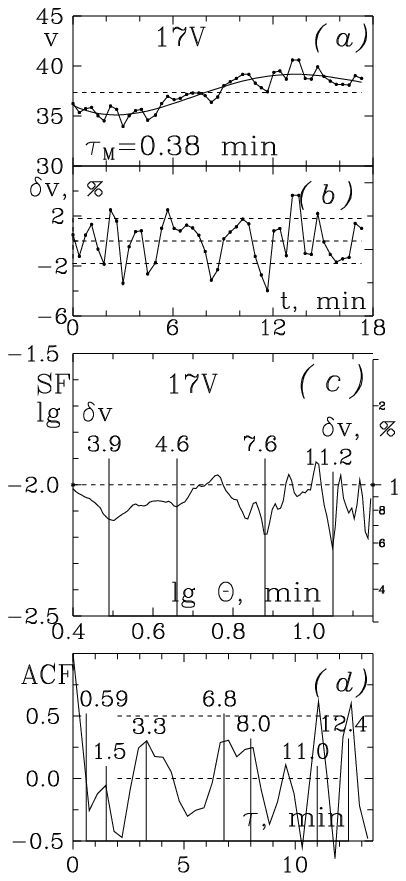, width=.49\textwidth}}
%\caption[] { }
\label{fig.2}   \end{center}  \end{figure*} % \vspace{-6mm}

\end{appendix}

\end{document}